\documentclass[aps,prapplied,floatfix]{revtex4-2}
\usepackage{graphicx}
\usepackage{amsmath}
\usepackage{siunitx}
\usepackage{hyperref}
\usepackage{subcaption}
\usepackage{chemformula}
\usepackage{mhchem}
\usepackage{comment}
\usepackage{textcomp}
\usepackage{float}
\DeclareUnicodeCharacter{2009}{\,}

\begin{document}

\title{Comparison of Two-Level System Microwave Losses in Pure Bulk Microcrystalline \ch{Nb2O5} and \ch{NbO2} Oxide Samples}

\author{Vishal Ganesan, Jiankun Zhang, Drew G. Wild, Alexey Bezryadin}
\affiliation{Department of Physics, University of Illinois at Urbana-Champaign, Urbana, IL 61801}

\begin{abstract}
Losses from two-level systems (TLS) associated with amorphous oxides remain one of the primary limitations to the performance of superconducting qubits and microwave cavities. Niobium resonators are widely used in quantum science experiments, yet niobium's natural oxide layer contains various types of oxides whose relative contributions to TLS loss have not been clearly distinguished. Here, we use a superconducting 3D microwave cavity to measure commercial 99.9\% trace metal pure, microcrystalline oxide powders \ch{Nb2O5} and \ch{NbO2} in bulk amounts. Using this approach, we directly compare the loss characteristics of \ch{Nb2O5} and \ch{NbO2}. Our measurements show that the nominal \ch{Nb2O5} bulk oxide powder samples exhibit losses which have the power and temperature behavior expected for TLS. Moreover, the measurements agree with existing theoretical models. Analogous measurements performed on \ch{NbO2} bulk powder samples do not show any detectable TLS loss signatures. Based on our results we propose that the TLS losses might be reduced if a high quality microcrystalline \ch{NbO2} oxide dominates the \ch{Nb2O5} oxide in practical Nb cavities. These results establish a materials based strategy for isolating oxide specific TLS losses and provide a reference measurement for niobium oxide phases relevant to superconducting quantum devices.
\end{abstract}

\maketitle

\section{Introduction}
Superconducting microwave resonators are essential components in quantum information processing, particle accelerator technology, and searches for new physics via dark matter experiments. Their performance is characterized by the intrinsic quality factor ($Q_\text{i}$), which limits photon lifetime and coherence \cite{Martinis2005, Krantz2019}. At low temperatures and excitation powers, dielectric losses arise from microscopic two-level systems (TLS) in amorphous materials, such as native oxides or dielectric interfaces. Measurements of $Q_\text{i}$ decreasing as microwave power is reduced is widely interpreted as a signature of these TLS losses \cite{Crowley2023,Vallieres2025}. TLS limit the achievable $Q_\text{i}$ and consequently the device performance \cite{TLS_overview, Takenaka2026, Romanenko2020}. Despite numerous mitigation strategies, including substrate removal \cite{Bruno2015}, oxide passivation \cite{deory2024low}, and surface treatments \cite{Kalboussi2024, Quintana2014, Chiaro2016}, TLS losses remain a persistent issue.

\par
For niobium, a common superconducting cavity material, TLS losses are generally attributed to its native oxide layers that form spontaneously in air. The niobium oxide system is arranged in a stack: metallic Nb covered by suboxides, followed by \ch{NbO} and \ch{NbO2}, and finally capped by \ch{Nb2O5} \cite{Halbritter1987, Murthy2022}. Although previous studies have suggested that the outer \ch{Nb2O5} layer dominates TLS losses \cite{Romanenko2017}, direct experimental verification has been hindered by the difficulty of isolating individual oxide phases. To distinguish their individual contributions, we purchase from a commercial supplier nominally pure oxide powders of \ch{Nb2O5} and \ch{NbO2}. We then use XRD (x-ray diffraction) to characterize them and find that \ch{Nb2O5} is monoclinic and \ch{NbO2} is tetragonal (see Appendix A for details). Then we employ a 3D superconducting \ch{Nb} cavity, and install in it a bulk amount of one oxide at a time 
(\ch{Nb2O5} powder or  \ch{NbO2} powder), and measure the quality factor versus the temperature or the microwave power. 

While we cannot rule out the possibility that defects within \ch{NbO} could, in principle, induce TLS‑like losses, we did not investigate \ch{NbO} powder in this study since we focus on insulating oxide phases. We note, however, that metallic or proximity‑effect–related mechanisms associated with \ch{NbO} have been discussed in literature in the context of microwave dissipation in other regimes \cite{Gurevich2017,Kubo2019}. Yet, any volume of \ch{NbO} oxide present as part of the thin native oxide layer on the cavity walls is negligible compared to the deliberately introduced bulk amounts of \ch{Nb2O5} or \ch{NbO2} oxide powder samples.

Thus, we measure the quality factor of the cavity with oxide powder samples versus temperature and microwave power for commercially purchased \ch{Nb2O5} or \ch{NbO2} powders. The results obtained on \ch{Nb2O5} powders are in agreement with the existing TLS models. At the same time, measurements on \ch{NbO2} powders do not show any dependence on microwave power within our measurement sensitivity. It is important to distinguish that the commercially purchased oxide powders differ from the natural oxide layer formed on \ch{Nb}, where the oxides in the natural layer are typically thought to be more defect-rich and possibly even amorphous. This is distinct from the oxides in this study, since, as we have determined by XRD (see Appendix A), the purchased \ch{Nb2O5} powder is largely microcrystalline with a monoclinic phase, and \ch{NbO2} is microcrystalline with a tetragonal phase. Therefore, we present direct evidence that only \ch{Nb2O5} is responsible for the TLS effect, while \ch{NbO2} is not for the samples in this study.

\section{Experimental Methods}

The oxide powders (\ch{Nb2O5} and \ch{NbO2}) were commercially obtained from Sigma-Aldrich with stated purity exceeding 99.9\% on a trace metal basis. 
XRD measurements confirm the manufacturer's purity claim (Appendix A). Approximately 50\,mg of powder was weighed and mixed with one drop ($\sim$200\,mg) of clear nail polish to form a uniform viscous suspension. The resulting mixture was applied to a 10\,mm\,$\times$\,10\,mm a-plane sapphire substrate, used to support the oxide load, with a clean toothpick producing a slug-shaped deposit at the center of the sapphire substrate. Sapphire provides a low-loss crystalline dielectric background, ensuring that observed dissipation originates primarily from the oxide sample. The sapphire substrates were ultrasonically cleaned in acetone and isopropanol (IPA) prior to oxide deposition. The resulting sample contained about 20\,mg of oxide. 
The nail polish acted as a mechanical binder and inert host that preserves the oxide microstructure and prevents particulate contamination of the cavity interior.

Both oxide powders follow the same preparation with analogous binder amounts and cavity conditions, yet our measurements show that only the \ch{Nb2O5} commercial bulk powder exhibits TLS losses, while the \ch{NbO2} powder samples do not show any measurable TLS effect in the power and temperature regimes studied. Specifically, our measurements show that $Q_\text{i}$ is reduced as microwave drive power is reduced for \ch{Nb2O5} samples, while we see an absence of such a reduction of $Q_\text{i}$ with reduced microwave power in \ch{NbO2} samples. (Note that we define, phenomenologically, "TLS loss" as an observation of $dQ_i/dP>0$ where $P$ is the microwave power.) We therefore conclude that the binder does not contribute to the TLS losses for the comparison of these oxides shown in this study.

\begin{figure}[H]
    \centering
    \includegraphics[width=0.95\linewidth]{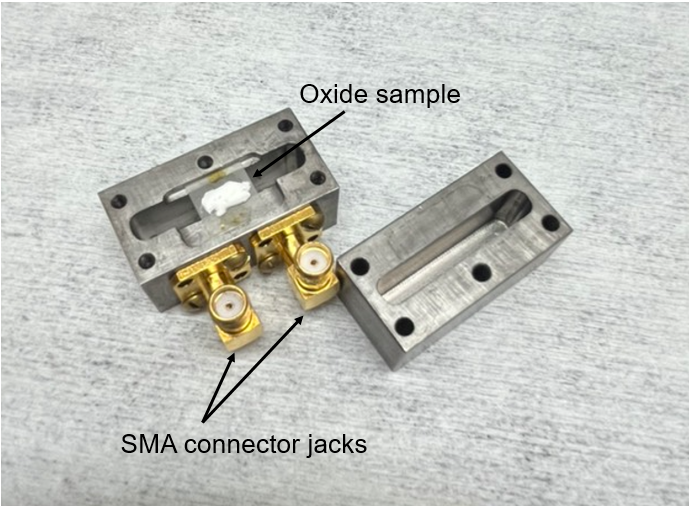}
    \caption{Sample Nb25\_NP5 on a sapphire substrate installed in the \ch{Nb} cavity. }
    \label{cavity}
\end{figure}

 The coated substrates were left to dry under ambient laboratory conditions for 24\,hours to ensure complete solvent evaporation. Once dried, the samples were mounted into a rectangular niobium cavity, with GE varnish securing the substrates within the machined groove at the electric field antinode for the fundamental resonant cavity mode, as shown in Fig.~\ref{cavity}. Samples are labeled by the oxide type (Nb02 or Nb25 for \ch{NbO2} and \ch{Nb2O5} respectively), mixture medium type (NP for nail polish), and version number. The one exception is the first sample created with a label "AP" standing for the cryogenic grease Apiezon (ex. Nb25\_NP3 indicates \ch{Nb2O5}, mixed with nail polish, being the 3rd sample made).

Before oxide sample installation, the \ch{Nb} cavity was cleaned by sonication in acetone and IPA for five minutes each. This is followed by a bake in a flow of \ch{Ar} for 1 hour at \SI{400}{\celsius} and 3\,Torr to remove surface contaminates and adsorbates. This annealing procedure is known to improve $Q_\text{i}$ \cite{Posen2020, Tamashevich2025}. Note that during sonication and annealing, the cavity SMA jack connectors are removed, and there is nothing installed in the groove of the cavity. Each experiment followed the nominally the same timeline. After annealing, the cavity was exposed to atmospheric conditions for 1 hour. This time includes transferring the cavity from the vacuum furnace to the sample installation area, and then subsequent installation in the cryostat. The cryostat is then sealed and pumped down from atmospheric pressure to about $10^{-6}$\,Torr in about another hour and then the cooling begins.

Therefore, the natural oxide layer that grows on the cavity walls during this time frame is similar for all our experiments. We also note that since our samples have a very large volume of commercially purchased oxide powder compared to the volume of the natural oxide layer grown when exposed to the atmosphere, we ignore this natural layer on the Nb cavity walls and treat it as a small perturbation. Additionally, our samples reside in the center of the cavity, where the electric field strength is the largest for the fundamental cavity mode, so again the contribution of the sample is dominant \cite{Pozar2012}.

The assembled cavity was anchored on the cooling stage of a He-3 cooler with a base temperature of $390$\,mK, or a dilution fridge with a base temperature of $60$\,mK. Transmission ($S_{21}$) measurements were performed using a Keysight P9374A vector network analyzer (VNA). VNA microwave power was varied to achieve an average power resonantly circulating, $P_\text{circ}$ (to be discussed below), in the cavity from about $-75$ to $+35$\,dBm (corresponding to a power resonantly applied at the cavity SMA connector jack, $P_{in}$, from $-110$ to $0$\,dBm). Note that the He-3 system was equipped with attenuators anchored at different stages of the cooling system. The total attenuation of the input line was $18$\,dB and the total attenuation of the output line was also $18$\,dB. There is an additional $+40$\,dB room temperature amplifier before the output signal reaches the VNA. The dilution fridge has $-60$\,dB attenuation on the input line, with a $+40$\,dB HEMT amplifier at the $4$\,K stage, and a $+40$\,dB room temperature amplifier. 

All data presented in Section IIIA, B and D are collected in the He-3 cooler with a base temperature of 390\,mK. The exception is the data shown in Section IIIC, where one \ch{Nb2O5} sample was installed in the dilution fridge with a base temperature of 60\,mK.

\begin{figure}[H]
\centering
\begin{subfigure}[b]{0.49\linewidth}
    \includegraphics[width=\linewidth]{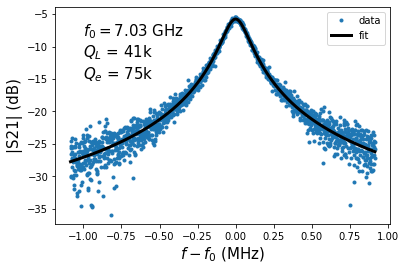}
    \caption{}
    \label{fig:empty_mag}
\end{subfigure}
\hfill
\begin{subfigure}[b]{0.49\linewidth}
    \includegraphics[width=\linewidth]{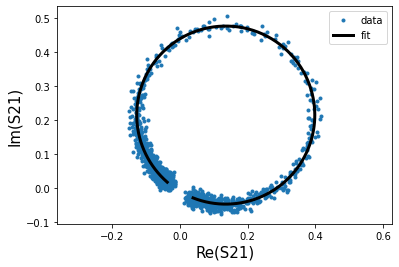}
    \caption{}
    \label{fig:empty_complex}
\end{subfigure}
\caption{Representative complex Lorentzian fit to the transmission coefficient of the cavity with the sample Nb25\_NP5 installed, at $P_\text{circ} = -45$\,dBm ($N\approx10^6$) at $390$\,mK. (a) This plot shows $S_{21}$ magnitude as a function of frequency. (b) This plot shows the corresponding complex-plane circle representation of S21 in the Argand plane. The fit to Eq.~\ref{eq:lorentzian} are shown by solid black lines. This plot is used to extract $Q_\text{i}$ and $Q_\text{e}$. Standard errors for the frequency and the quality factors obtained from this circular fit are several orders of magnitude below the mean values for these quantities.}
\label{exampleFit}
\end{figure}

The transmission $S_{21}$ near resonance is modeled as a complex Lorenztian given by \cite{Chen2022a,Chen2022b}

\begin{equation}
S_{21}(f) = \frac{(Q_\text{L}/|Q_\text{e}|)e^{i\phi}} {1 + 2iQ_\text{L}\frac{f - f_0}{f_0}},
\label{eq:lorentzian}
\end{equation}

where $f$ is the frequency, $f_0$ is the resonance frequency, $\phi$ accounts for impedance mismatch, $Q_\text{L}$ is the loaded quality factor, $Q_\text{e}=|Q_\text{e}|e^{-i\phi}$ is the external coupling quality factor, and the intrinsic quality factor $Q_\text{i}$ is obtained from the fit parameters $Q_\text{L}$ and $Q_\text{e}$ using the formula $\frac{1}{Q_\text{i}} = \frac{1}{Q_\text{L}} - \frac{1}{|Q_\text{e}|}$. Each resonance was calibrated by removing (mathematically) the effects of line attenuation and amplification, then fit with Eq.~\ref{eq:lorentzian} to both the real and imaginary components of $S_{21}$ (see Fig.~\ref{exampleFit}) at each power applied by the VNA. For low powers, many scans are averaged to improve the signal-to-noise ratio. This circular fit allows us to obtain accurate values for the intrinsic quality factor.

Eq. \ref{eq:lorentzian} follows the standard two-port model used for transmission-type resonators. This form differs from the absorption-type expression commonly used in the superconducting qubit literature, where the resonator is coupled to a transmission line, and the transmission takes the form 
$S_{21}(f) = 1-\frac{(Q_\text{L}/|Q_\text{e}|) e^{i\phi}}{1 + 2iQ_\text{L}\frac{f - f_0}{f_0}}$ \cite{Khalil2012,Probst2015}. The difference arises solely from the coupling geometry. Our cavity is measured in a through-transmission geometry, so the baseline transmission does not include the unity term. Both forms are mathematically equivalent under the appropriate coupling configuration.

\section{Results}
\subsection{Control Measurements}

\begin{figure}[H]
    \centering
    \includegraphics[width=0.95\linewidth]{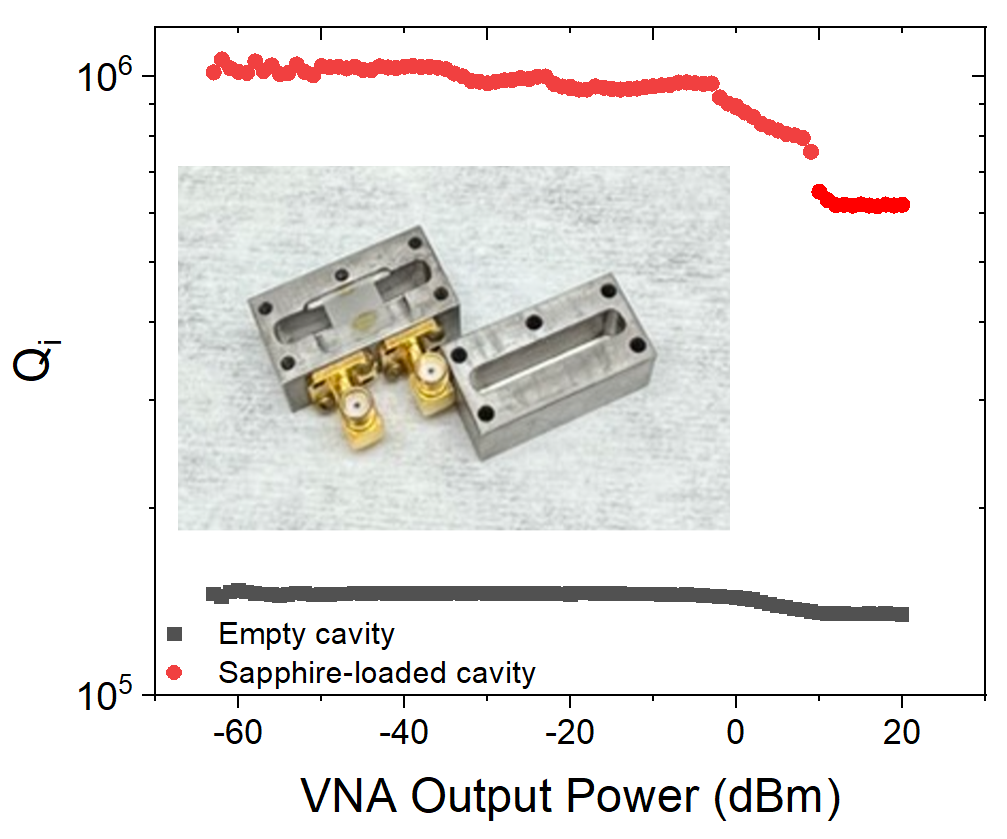}
    \caption{Control measurements of $Q_\text{i}$ versus VNA output power of the empty \ch{Nb} cavity (black) and the cavity loaded with the bare, clean sapphire substrate (red) at $390$\,mK. In these measurements there were no \ch{Nb} oxide sample. The inset shows the cavity with the bare, clean sapphire substrate. }
    \label{control}
\end{figure}

In Fig.\ref{control} we show control measurements of $Q_\text{i}$ versus VNA output power for the empty cavity (
nothing installed in the groove
) and the cavity loaded with a bare, clean sapphire substrate in the groove (no oxide sample on the substrate). Before installation of the empty cavity or the sapphire-loaded cavity in the He-3 cooler, the cavity underwent the cleaning procedure described above. The figure shows the intrinsic quality factor, which was of the order of 200\,k for the empty cavity  and of the order of 1\,M for the cavity loaded with an bare, clean sapphire substrate. Note that at low microwave power the $Q_\text{i}$ does not go down. Therefore, these control measurements indicate that a pronounced TLS-like signature is not observed within the explored power range and sensitivity of our baseline cavity, where other residual-loss mechanisms (e.g., surface-related losses including possible subgap state contributions \cite{Kubo2022, Gurevich2023} and/or trapped flux \cite{Posen2016, Huang2016}) may dominate losses at low temperatures.

Surprisingly, we find that loading the cavity with a bare, clean sapphire substrate dramatically improves the quality factor compared to the empty cavity with nothing installed in the groove. We attribute this to the increased concentration of the electric field around the substrate when it is introduced. This reduces electric field interactions with imperfectly machined walls or any native oxide on the walls. Both these examples may reduce the quality factor, but do not produce TLS-like losses. Further studies of this effect is the subject of future work.

\subsection{\ch{Nb2O5} Power Measurements}

\begin{figure}[H]
    \centering
    \includegraphics[width=0.95\linewidth]{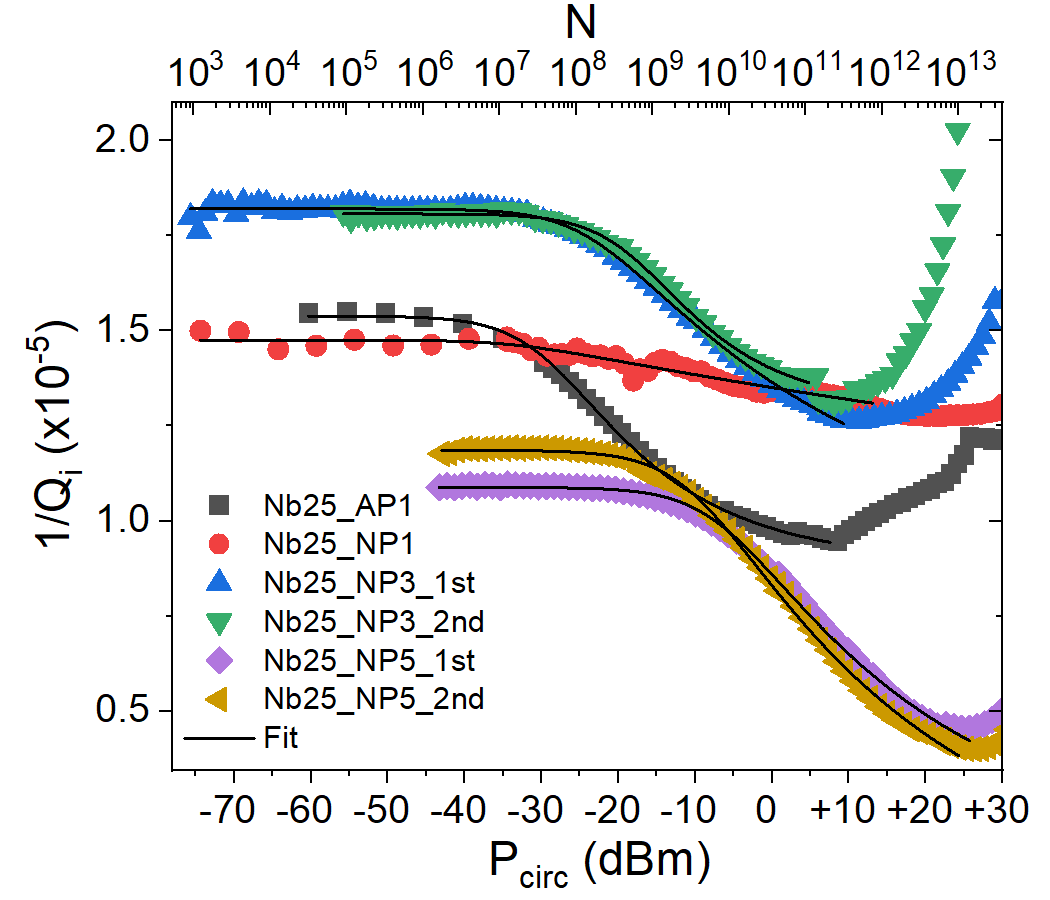}
    \caption{Power dependence of the inverse of the intrinsic quality factor $Q_\text{i}$ for \ch{Nb2O5} samples as a function of circulating power (bottom axis) and photon number N (top axis) at $390$\,mK. Solid black lines show the TLS saturation model fit of Eq.~(\ref{eq:TLSmodel}). Two samples, Nb25\_NP3 and Nb25\_NP5, were measured twice (indicated by $1^{st}$ and $2^{nd}$) to confirm TLS losses are reproducible after removing the samples, cleaning the cavity, and re-installing the samples. }
    \label{Nb2O5_TLS}
\end{figure}

In Fig.~\ref{Nb2O5_TLS} we show measurements of \ch{Nb2O5} bulk samples that exhibit a clear reduction in $Q_\text{i}$ at low circulating powers. All samples are nominally the same. Such a drop off of the quality factor at low power is similar to the previously observed behavior of oxidized Nb resonators due to TLS losses \cite{Romanenko2017, Lindstrom2009}.

For TLS-dominated losses, the intrinsic quality factor can be modeled as \cite{Gao2008, TLS_glasses}:

\begin{equation}
\frac{1}{Q_\text{i}(P_\text{circ})} = \frac{F \tan\delta_\text{TLS}}{(1 + P_\text{circ}/P_\text{c})^\beta} \tanh{\left(\frac{h f_0}{2 k_B T}\right)} + \frac{1}{Q_\text{other}}
\label{eq:TLSmodel}
\end{equation}
This formula was derived from the standard tunneling model for non-interacting, resonantly coupled, TLS. Here $F$ is the participation ratio, which is the fraction of the electric field energy in the sample divided by the total electric field energy in the cavity, $\tan\delta_\text{TLS}$ is the loss tangent of the TLS hosting material, $P_\text{c}$ is the critical circulating power at which TLS are saturated, $\beta$ is a phenomenological exponent, $T$ is the temperature, $k_B$ is Boltzmann's constant, $h$ is Planck's constant, and $Q_\text{other}$ accounts for residual non-TLS losses which become appreciable at high power and temperature (such as thermally activated quasiparticle losses). The power resonantly circulating in the cavity is $P_{circ} =\frac{P_{in}Q_\text{L}^2}{\pi Q_\text{e}}$, where the average total photon number inside the cavity is found as
$N = \frac{P_\text{circ}}{h f_0^2}$  \cite{Goetz2016}.

In our geometry, the intentionally introduced oxide powder sample is by far the dominant source of dielectric loss. The native oxide on the cavity walls is only a few nanometers thick and contributes negligibly compared to the milligram‑scale sample placed directly at the electric‑field antinode. As a result, the resonance characteristics, namely the observed saturation behavior, are governed almost entirely by the sample rather than by the background cavity surfaces. Although the TLS saturation model is formally expressed in terms of the local electric field within the dielectric material, in this configuration the relevant field is the one experienced by the sample, and this field scales monotonically with the applied circulating power. We therefore use $P_{circ}$ as representative quantity for the local electric field at the sample location, with the fitted critical power $P_c$ reflecting the response of the sample rather than the cavity walls.

The solid black lines in Fig.~\ref{Nb2O5_TLS} show the fits to Eq.~\ref{eq:TLSmodel}, which agree well with the data. Fit parameters are listed in Table.~\ref{Nb25-fitresults} and the fits we have done on all our samples give out a very small $\beta$ term as compared to the standard tunneling model used to derive Eq.~\ref{eq:TLSmodel}, which predicts $\beta = 0.5$ for non-interacting TLS. Previous studies report fits to the power dependence that also require a small $\beta$ term. This reduction of beta has been interpreted as evidence for interacting TLS ensembles \cite{Burnett2014,Faoro2015, TLS_overview}. This supports the interpretation that \ch{Nb2O5} hosts an ensemble of interacting TLS that are the cause of the observed loss at low power.

%The critical power $P_c$ is observed to systemically increase in later samples, i.e., those at the bottom of Table~\ref{Nb25-fitresults}, as the cavity cleaning and annealing procedures were refined. This trend is consistent with previous reports showing that surface treatments such as high-temperature annealing and solvent-based cleaning effectively suppress extrinsic (non-TLS) loss contributions associated with adsorbates or hydrocarbon residues on cavity walls \cite{Megrant2012, Romanenko2017, TLS_overview}. 

The \ch{Nb2O5} powder samples in our experiment do not contain any metallic \ch{Nb}. (And, obviously, the oxide sample is not in physical contact with the superconducting niobium cavity walls.) Therefore these results demonstrate that the TLS loss originates within the bulk of the oxide, or, possibly, at the oxide-vacuum interface, rather than at any metal-oxide or metal-vacuum interface. This observation provides strong evidence that the dominant TLS in the present configuration are intrinsic to the amorphous \ch{Nb2O5}, in agreement with previous works identifying oxygen vacancy-related dipoles as the primary source of dielectric loss in niobium cavities \cite{Bafia2024, DuBois2013}.

\begin{table}[H]
\centering
\caption{Fit parameters extracted from Eq.~\ref{eq:TLSmodel} for various \ch{Nb2O5} samples. Standard errors are obtained from fits, and errors for $P_\text{c}$ were omitted for clarity, but are 1 order of magnitude below reported values. $F \tan\delta_\text{TLS}$ is approximated as $F \delta_\text{TLS}$, and $\delta_\text{TLS}$ is abbreviated as just $\delta$.}
\begin{ruledtabular}
\begin{tabular}{c c c c c}
Sample & $ F\delta \: (\times10^{-5})$ & $\beta\: (\times10^{-1})$  &  $P_c$ (\,dBm)  & $\delta  \: (\times10^{-4})$ \\
\hline
Nb25\_AP1                    & $1.3\pm0.2$   & $2.9 \pm 0.1$    & $-31.8$     &$6.5$\\
Nb25\_NP1                    & $1.2\pm0.1$   & $0.3 \pm 0.1$    & $-35.5$    &$6.1$\\
Nb25\_NP3\_1$^{\mathrm{st}}$ & $1.8\pm0.1$   & $1.2 \pm 0.1$    & $-26.0$     &$9.2$\\
Nb25\_NP3\_2$^{\mathrm{nd}}$ & $1.0\pm0.4$   & $3.8 \pm 0.4$    & $-20.1$     &$5.1$\\
Nb25\_NP5\_1$^{\mathrm{st}}$ & $1.9\pm0.1$   & $1.5 \pm 0.1$    &  $-9.0$    &$9.7$\\
Nb25\_NP5\_2$^{\mathrm{nd}}$ & $2.3\pm0.2$   & $1.5 \pm 0.1$    &  $-12.6$    &$12$\\
\end{tabular}
\label{Nb25-fitresults}
\end{ruledtabular}

\end{table}

High frequency electromagnetic simulations using ANSYS HFSS were used to determine the participation ratio (see Appendix B for details), which yields $F \approx 0.02$. This leads to an average loss tangent $\delta = 8.1\times10^{-4}$ for the samples reported in Table \ref{Nb25-fitresults}, consistent with previous works \cite{Burnett2017, Verjauw2021, Kalboussi2025}. 

\subsection{\ch{Nb2O5} Temperature Measurements}
\begin{figure}[H]
\centering
\begin{subfigure}[b]{0.35\linewidth}
    \includegraphics[width=\linewidth]{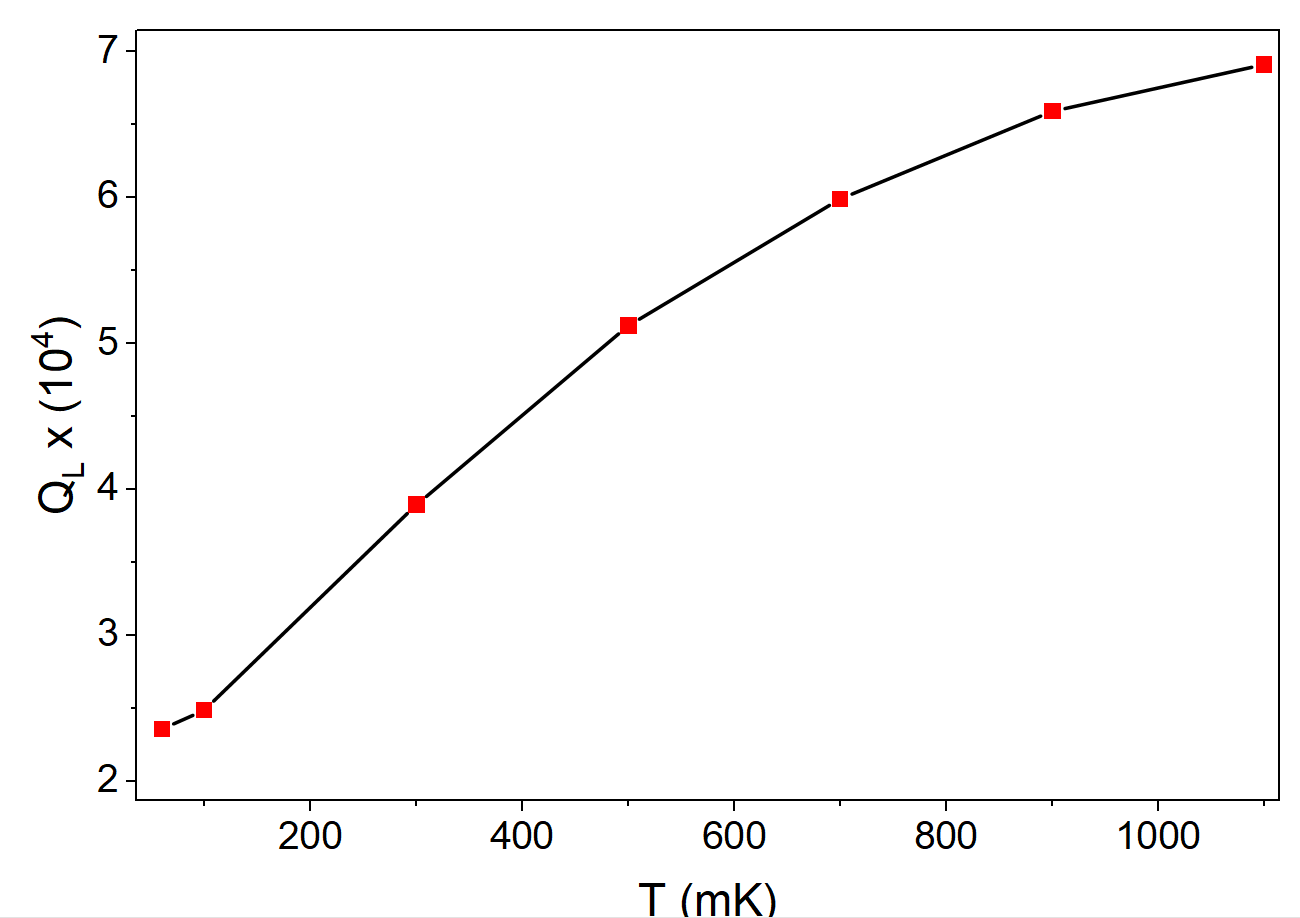}
    \caption{}
    \label{MrFrzQLRaw}
\end{subfigure}
\bigskip
\begin{subfigure}[b]{0.35\linewidth}
    \includegraphics[width=\linewidth]{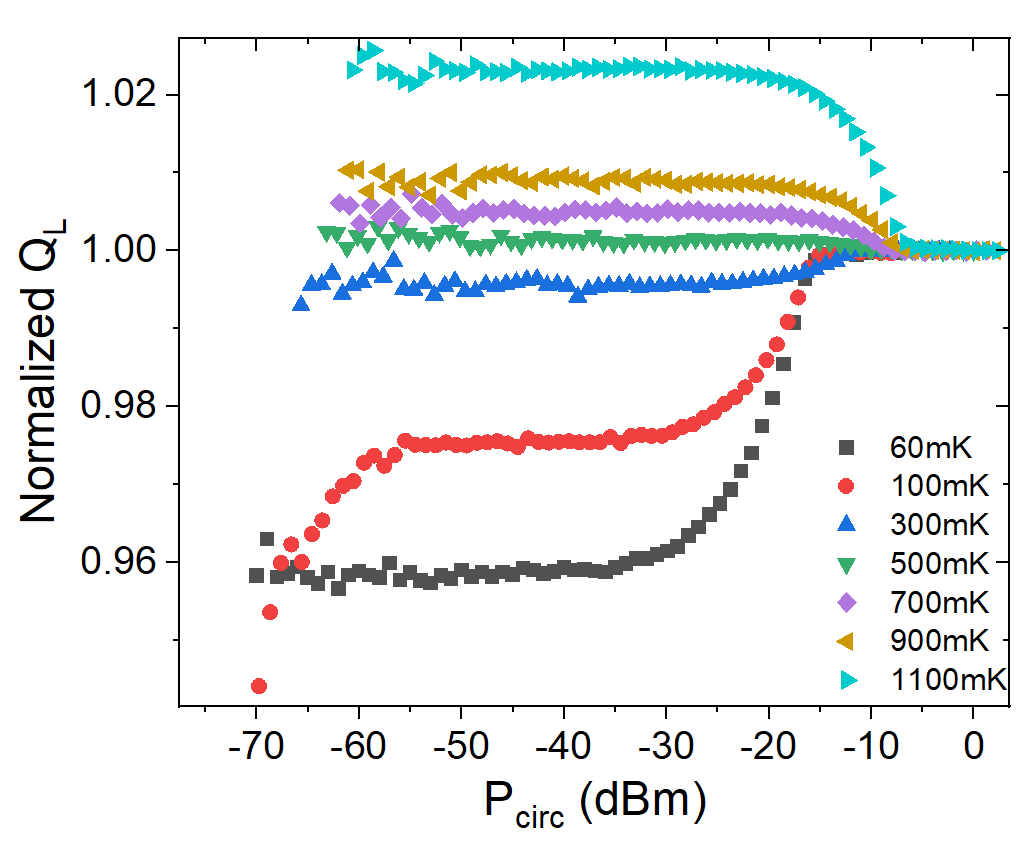}
    \caption{}
    \label{MrFrzQL}
\end{subfigure}
\bigskip
\begin{subfigure}[b]{0.6\linewidth}
    \includegraphics[width=\linewidth]{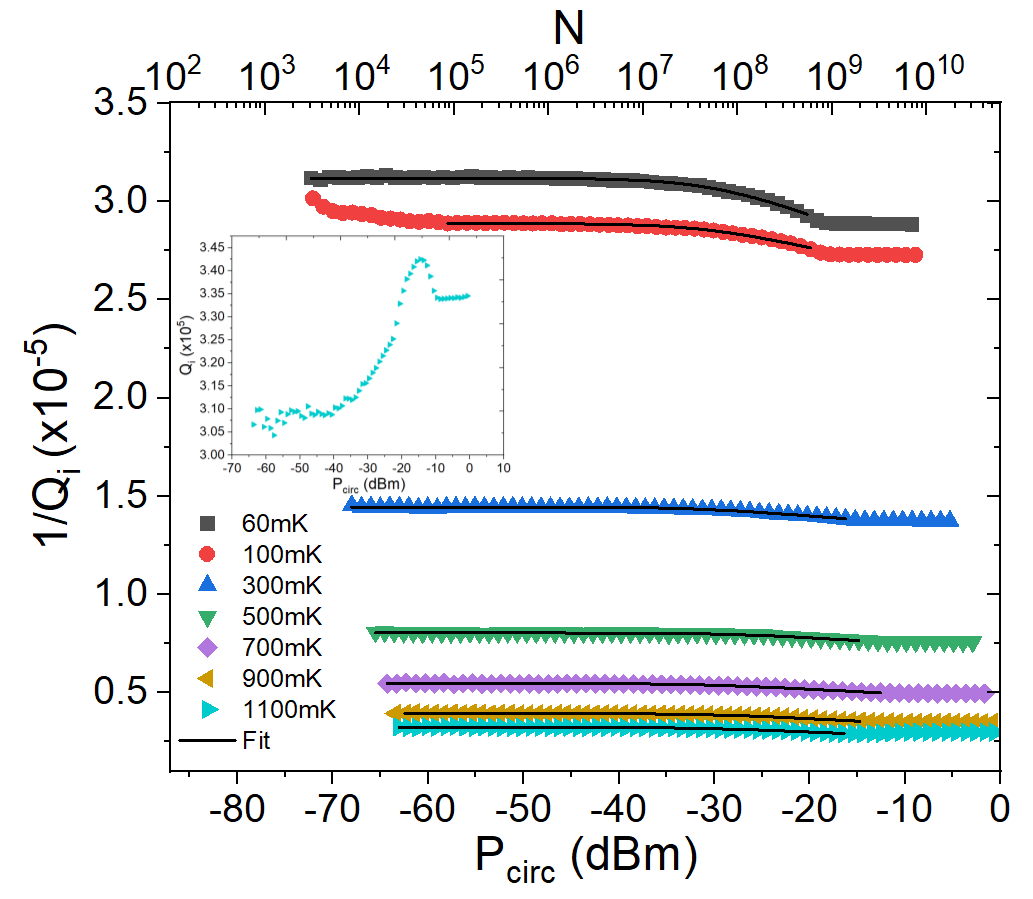}
    \caption{}
    \label{MrFrzQi}
\end{subfigure}
\caption{Sample Nb25\_NP2. (a) Measured loaded quality factor $Q_\text{L}$ as a function of temperature, at $P_\text{circ}=-60$\,dBm ($N\approx10^5$). The black line connects data points. (b) Loaded quality factor $Q_\text{L}$, normalized to its value at $P_\text{circ}=-10$\,dBm ($N\approx10^{10})$, as a function of circulating power for various temperatures.  (c) Inverse $Q_\text{i}$ versus circulating power (bottom axis) and photon number (top axis) with fits to Eq.~\ref{eq:TLSmodel} shown as solid black curves at various temperatures. The inset highlights a pronounced TLS-induced reduction in $Q_\text{i}$, measured at $1.1$\,K as a function of circulating power. This is in contrast to the absence of an obvious corresponding decrease in $Q_\text{L}$.}
\label{MrFrzGraph}
\end{figure}

To investigate losses associated with \ch{Nb2O5} as a function of temperature, the sample Nb25\_NP2 was measured in a dilution refrigerator with a base temperature of $60$\,mK. The empty cavity underwent the cleaning procedure described above. In Fig.~\ref{MrFrzQLRaw} we show the raw data for the measured quality factor, taken at a low circulating power of $-60$\,dBm ($N\approx10^5$). It is clear that the loaded quality factor drops drastically at low temperature, giving evidence of TLS losses. In Fig.~\ref{MrFrzQL} we show the normalized loaded quality factor $Q_\text{L}$ at various temperatures, showing clear TLS-like loss behavior below $500$\,mK. However above $500$\,mK, $Q_\text{L}$ exhibits no clear signature of TLS-related loss at low drive power. In contrast, $Q_\text{i}$ is reduced at low powers for all temperatures (Fig.~\ref{MrFrzQi}), as illustrated in particular by the measurement at $1.1$\,K in the inset of Fig.~\ref{MrFrzQi}. 

We attribute this obvious absence of TLS losses in the measured $Q_\text{L}$ above $500$\,mK due to increased thermal quasiparticle losses resulting in increased surface resistance and non-linear effects, so losses may be primarily dominated by these effects. Moreover, a larger fraction of TLS are thermally activated and cannot readily absorb microwave photons, so the TLS population available to induce losses is decreased, and therefore an obvious decrease in $Q_\text{L}$ is less apparent \cite{Kubo2025,Dhakal2024}. Thus we see the presence of TLS losses at all temperatures measured, given that $Q_\text{i}$ decreases as $P_{circ}$ decreases. 

A pronounced result is that $Q_\text{i}$ increases strongly as the temperature increases. In other words $Q_\text{i}$ at $1.1$\,K is about 10 times larger than $Q_\text{i}$ at $60$\,mK (at $P_\text{circ} = -45$\,dBm). Such a strong effect has not been observed previously on oxidized Nb cavities. This fact supports a model in which TLS are thermally activated and weakly coupled to a thermal bath \cite{Goetz2016, Bruno2015}.

In Fig.~\ref{MrFrzQi}, the solid black curves show the fits to Eq.~\ref{eq:TLSmodel}, which agree well with the data. Fitted parameters are listed in Table~\ref{MrFrzFit} and show no significant variation in the critical power $P_\text{c}$ or the exponent $\beta$ across the measured temperature range. We find that the $\beta$ terms at all temperatures are smaller or even much smaller than the expected $0.5$ value, which \textbf{again} suggests TLS-TLS interactions in our \ch{Nb2O5} bulk samples. As a reminder, we note that the exponents extracted from the power dependence measurements presented in Table~\ref{Nb25-fitresults} also led to the same conclusion.

Notably at 100\,mK, $Q_\text{i}$ shows a pronounced drop which begins near $P_\text{circ} = -20$\,dBm, after which it settles into a plateau at a value of approximately $3.2 \times 10^4$ that continues to $-60$\,dBm (see Fig.\ref{MrFrzGraph}). Our interpretation is as follows. Between $P_\text{circ} = -20$ to $-30$\,dBm TLS induced losses increase as power is lowered, since a larger fraction of TLS remain unsaturated and therefore able to absorb resonant photons. Because each TLS can absorb only one photon, the loss saturates since the majority of active TLS are engaged, leading to a nearly constant $Q_\text{i}$ from $P_\text{circ} = -30$ down to $-60$\,dBm. Interestingly below $-60$\,dBm, $Q_\text{i}$ decreases again, a feature not seen for other temperatures. This unexpected secondary loss process is very unusual. We speculate that it may suggest multiple and distinct TLS populations within this oxide. A more detailed investigation of this behavior will be the subject of future work

%We plan a detailed investigation of this phenomenon in future research projects.

%but it may indicate the presence of multiple, distinct TLS populations.

%This unexpected secondary loss process was not explored further, but may suggest multiple and distinct TLS populations.

\begin{table}[H]
\centering
\caption{Fit parameters for sample Nb25\_NP2 at various temperatures. Standard errors obtained from fits to Eq.~\ref{eq:TLSmodel} are 2 orders of magnitude below reported values, and are omitted for clarity.}
\begin{tabular}{c c c c c}
\hline
\hline
$T$ (\,mK) & $F\delta \: (\times10^{-5})$ & $\beta (\times10^{-2})$ & $P_c$ (\,dBm)  & $\delta\: (\times10^{-3})$\\
\hline
60   & $3.2$     & $2.6$           & $-29.5$   &$1.6$\\
100  & $3.1$     & $1.7$           & $-30.8$  &$1.6$\\
300  & $2.8$     & $1.2$           & $-31.3$   &$1.4$\\
500  & $2.5$     & $1.8$           & $-37.1$  &$1.2$\\
700  & $2.3$     & $2.2$           & $-30.7$   &$1.2$\\
900  & $2.1$     & $3.0$           & $-29.2$  &$1.1$\\
1100 & $2.1$     & $3.0$           & $-30.7$  &$1.1$\\
\hline
\hline
\end{tabular}
\label{MrFrzFit}
\end{table}

\begin{figure}[H]
    \centering
    \includegraphics[width=0.95\linewidth]{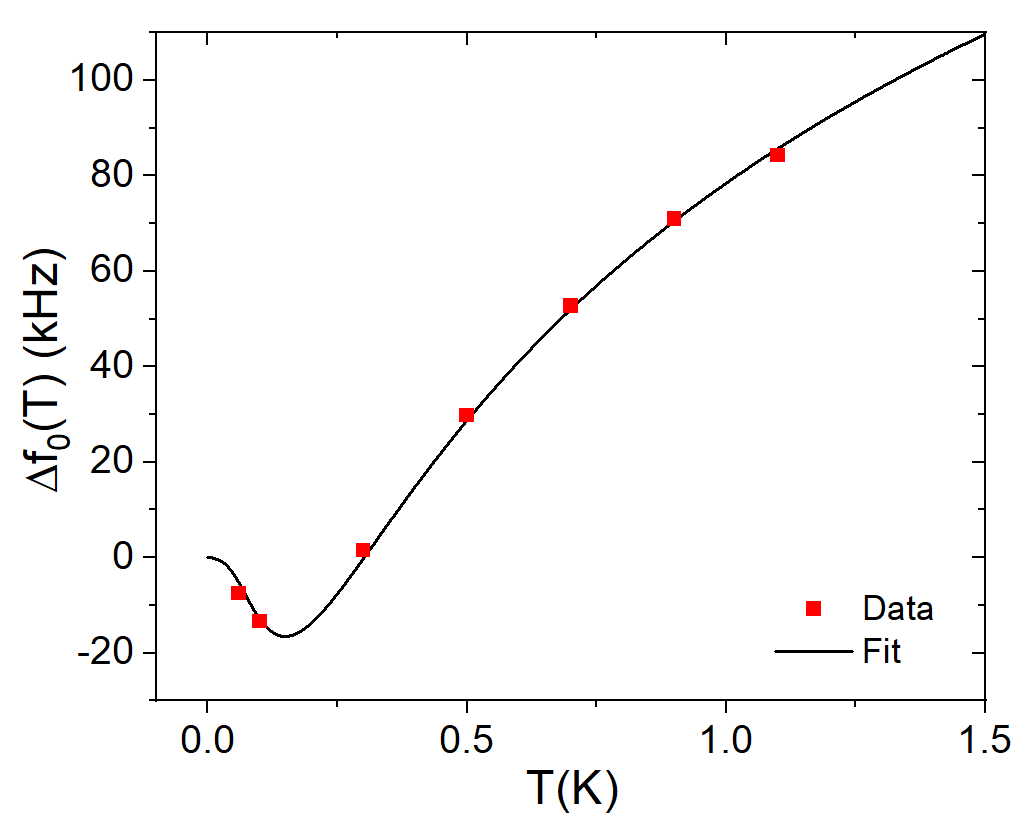}
    \caption{Sample Nb25\_NP2. Measured resonance frequency shift (red squares) at $P_\text{circ}=-45$\,dBm ($N\approx10^6$), as a function of temperature. The solid black curve shows the fit to Eq.\ref{DiGamma}, extrapolated to wider range outside the measured temperature range.}
    \label{DeltaF}
\end{figure}

In Fig.~\ref{DeltaF} we show the measured shift of resonance frequency arising from non-resonant, dispersive TLS as a function of temperature, at constant $P_\text{circ}=-45$\,dBm ($N\approx10^6$), for sample Nb25\_NP2. These {\it non-resonantly} coupled TLS do not cause additional dissipation, but instead modify the real part of the dielectric response causing a measurable shift in the resonant frequency. On the qualitative level, the effect is analogous to the well known dispersive shift of the resonance of a microwave cavity due to a coupled qubit. Note that, in relation to our previous discussion, the effect on $Q_\text{i}$ by \textit{resonantly-coupled} TLS described in Eq.~\ref{eq:TLSmodel} reflects the corresponding modification for the imaginary component of the dielectric response \cite{Pappas2011, Wisbey2010}.

The temperature dependent resonant frequency shift is modeled as \cite{Gao2008}

\begin{equation}
\begin{split}
\Delta f_0(T) = 
\frac{F\delta}{\pi} \Bigg[
&\text{Re} \left(\Psi\!\left( 
      \frac{1}{2} + \frac{1}{2\pi i}\frac{h f_0(T)}{k_B T}
   \right)\right)  \\
&- \ln\!\left(
      \frac{1}{2\pi}\frac{h f_0(T)}{k_B T}
   \right)
\Bigg] f_0(0)
\end{split}
\label{DiGamma}
\end{equation}

Where $\Delta f_0(T) = f_0(T) - f_0(0)$ is the shift relative to zero-temperature resonance frequency, and $\text{Re}(\Psi)$ is the real part of the digamma function. This expression is found with the same theoretical basis used to find Eq.~\ref{eq:TLSmodel}, providing an alternative way to determine $\delta$. The solid black line in Fig.~\ref{DeltaF} shows the fit to Eq.~\ref{DiGamma}, which agrees well with the data. We find that $\delta = 1.8\times10^{-3}$  (using $F \approx 0.02$), which compares well with values reported in Table~\ref{Nb25-fitresults} and ~\ref{MrFrzFit}.

Note that the value of $\delta$ extracted from Eq.~\ref{eq:TLSmodel} represents a local TLS loss tangent, sampled only within a narrow bandwidth around resonance frequency $f_0$ (we call it Method I). In contrast, the value obtained from Eq.~\ref{DiGamma} corresponds to an average loss tangent over a much broader range of TLS excitation frequencies (Method II). Because the TLS density of states can vary outside the resonator bandwidth (vary with TLS excitation energy), the two methods generally yield slightly different values of $\delta$. Method I yields $\delta = 8.1\times10^{-4}$, and method II yields $\delta = 1.8\times10^{-3}$. Our results follow this expected trend and are consistent with prior studies, which report that Method II typically produces a $\sim$20\% larger loss tangent \cite{Pappas2011, Bruno2015}.

\subsection{\ch{NbO2} Power Dependence Measurements}
\begin{figure}[H]
    \centering
    \includegraphics[width=0.95\linewidth]{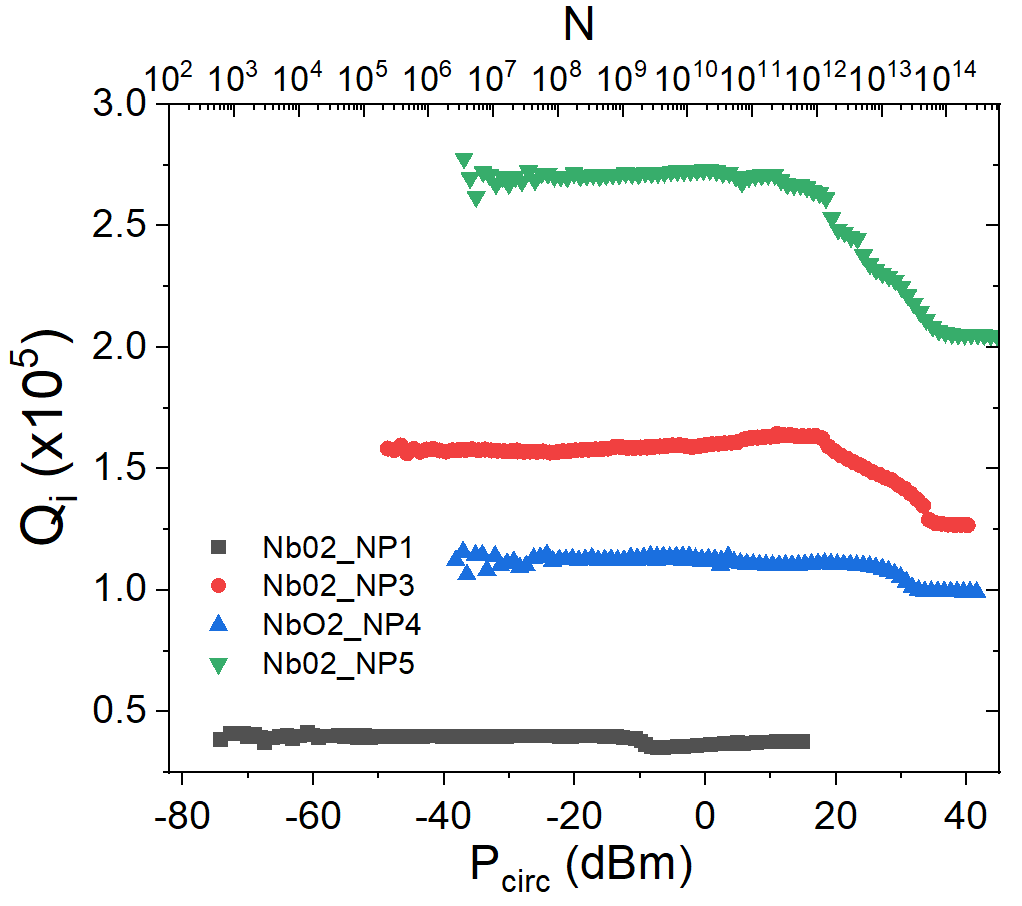}
    \caption{$Q_\text{i}$ versus circulating power (bottom axis) and photon number (top axis) for \ch{NbO2} bulk samples at $390$\,mK. The absence of a clear drop of $Q_i$ at low powers indicates that the \ch{NbO2} samples do not contribute to TLS losses for within our measurement sensitivity limits.}
    \label{NbO2_NoTLS}
\end{figure}

In Fig.~\ref{NbO2_NoTLS}, we show that \ch{NbO2} samples show no measurable TLS-like power dependence, where $Q_\text{i}$ remains constant over several orders of magnitude in circulating power. The \ch{NbO2} bulk samples are all nominally the same. The sample \text{Nb02\text{\_}NP1} has a very low quality factor due to the cavity being not diligently cleaned. This might also explain why there is a non-monotonic behavior in $Q_\text{i}$ manifested by the dip around $P_\text{circ} =-10$\,dBm ($N\approx10^9$). Importantly, these measurements also indicate that the nail polish used to bind the oxide powders is not a source of TLS. 

For all samples, we measure a pronounced decrease in $Q_\text{i}$ at high circulating power, followed by a plateau. We attribute this observed drop in $Q_\text{i}$ at higher circulating powers to the increased generation of nonequilibrium quasiparticles\cite{Glazman2021}, which are known to increase dissipation\cite{Barends2011, Liu2024} and reduce the quality factor in microwave resonators at high circulating powers \cite{Kubo2025,Goldie2013}. Prior studies have shown that high microwave powers can create quasiparticles through pair-breaking processes \cite{deVisser2014, Liu2024}, which become stronger as the applied power is increased. However, this does not explain the observed plateau at even higher circulating powers in nearly all datasets reported.

To explain both the drop of the $Q_\text{i}$ at higher circulating power and the observed plateau, we can follow Ref. \cite{Sung2026} and assume that a thin layer of \ch{Nb} hydride is present on the walls of the cavity and acts as a proximity-coupled superconductor \cite{Sung2024,Romanenko2013d}. At higher circulating powers, this thin layer of \ch{Nb} hydride begins to become a normal conductor, and so a large population of quasiparticles or normal electrons is released within the cavity walls. Therefore a drop in $Q_\text{i}$ occurs. At a characteristic power, this surface inclusions of niobium hydride are fully driven normal and the corresponding energy loss becomes power‑independent, so further increases in circulating power no longer reduce $Q_\text{i}$, producing the observed plateau.  Such an effect where $Q_\text{i}$ drops as power increases is observed frequently in superconducting radio-frequency cavity studies \cite{Romanenko2013,Romanenko2017, Romanenko2013d, Romanenko2020, Dhakal2024, Martinello2018}, but the plateau has not been previously observed.

\section{Discussion and Conclusions}

These results clearly identify \ch{Nb2O5} samples in this study as a principal host of TLS, while the \ch{NbO2} samples contribute to TLS losses negligibly under analogous experimental conditions. It was previously suggested that TLS in \ch{Nb2O5} arise from intrinsic structural disorder that can manifest as dangling bonds or oxygen vacancies that create under-coordinated Nb sites \cite{DuBois2013}. These sites can reorient between metastable configurations, forming electric dipoles that couple to the cavity field and yield the observed TLS behavior, where $Q_\text{i}$ drops off as circulating power decreases and then remains constant at even lower powers.

The contrasting TLS behavior observed between the two oxides is consistent with their underlying crystal structures. The \ch{Nb2O5} samples are monoclinic, which is known to have low symmetry and may host more defects in the crystal structure. This in turn can host many TLS candidates, such as oxygen vacancies or dangling bonds. Meanwhile, the \ch{NbO2} samples are tetragonal, which are much more symmetric and therefore will not host as many defects in the crystal structure \cite{Kittel2004, Tilley2008, Hammond2001}.  The more constrained structure of tetragonal crystals limit the number of defect configurations that can form bistable dipoles capable of coupling to the microwave field. Consequently, the monoclinic \ch{Nb2O5} samples exhibit clear TLS‑related loss, whereas the tetragonal \ch{NbO2} samples do not.

 %\begin{comment}This may explain prior observations that removing or thinning the outer \ch{Nb2O5} layer, %whether by HF etching or high-temperature annealing, significantly improves superconducting resonator quality %factors \cite{Romanenko2017, Posen2020, Romanenko2020}. Even when other native oxides such as \ch{NbO2} %remain, eliminating the defect rich \ch{Nb2O5} layer is sufficient to suppress TLS losses.\end{comment}

We model the electric field distribution in the cavity and the sample using ANSYSS HFSS and thus obtain the participation ration. This leads us to the conclusion that the loss tangent for the \ch{Nb2O5} is in the range $10^{-4}<\delta<10^{-2}$, which is in agreement with previous publications \cite{Burnett2017, Verjauw2021}. In particular the natural oxide occurring in a 3D microwave cavity, measured in \cite{Kalboussi2025}, showed a similar value, namely $10^{-4}<\delta<10^{-3}$.

In conclusion, we have developed a cavity-based method to isolate dielectric TLS losses associated with distinct, commercial bulk microcrystalline \ch{Nb} oxide phases. We demonstrate that the \ch{Nb2O5} oxide powder exhibits TLS-dominated losses, while \ch{NbO2} does not. Fits to the standard TLS model indicate TLS-TLS interactions in \ch{Nb2O5}, evidenced by the small $\beta$ term obtained for all measurements. Since our samples do not contain any Nb-Nb oxide interface, the results prove that TLS can originate from the Nb oxide itself. This approach provides a robust framework for studying oxide-specific dissipation mechanisms in microwave settings and offers a chemically selective, quantitative means to probe such effects.

\section{Acknowledgments}
The work was supported in part by the NSF DMR-2104757 and by the NSF OMA 2016136 Quantum Leap Institute for Hybrid Quantum Architectures and Networks (HQAN). This research was carried out in part in the Materials Research Laboratory Central Research Facilities, University of Illinois.

\section{Author Declarations}

The authors claim no conflicts of interest. The data that support the findings of this study are available upon reasonable request.

\bibliographystyle{unsrt}
\bibliography{references}   

\clearpage

\appendix

\section{X-ray Diffraction}

\begin{figure}[H]
\centering
\begin{subfigure}[b]{0.45\linewidth}
    \includegraphics[width=\linewidth]{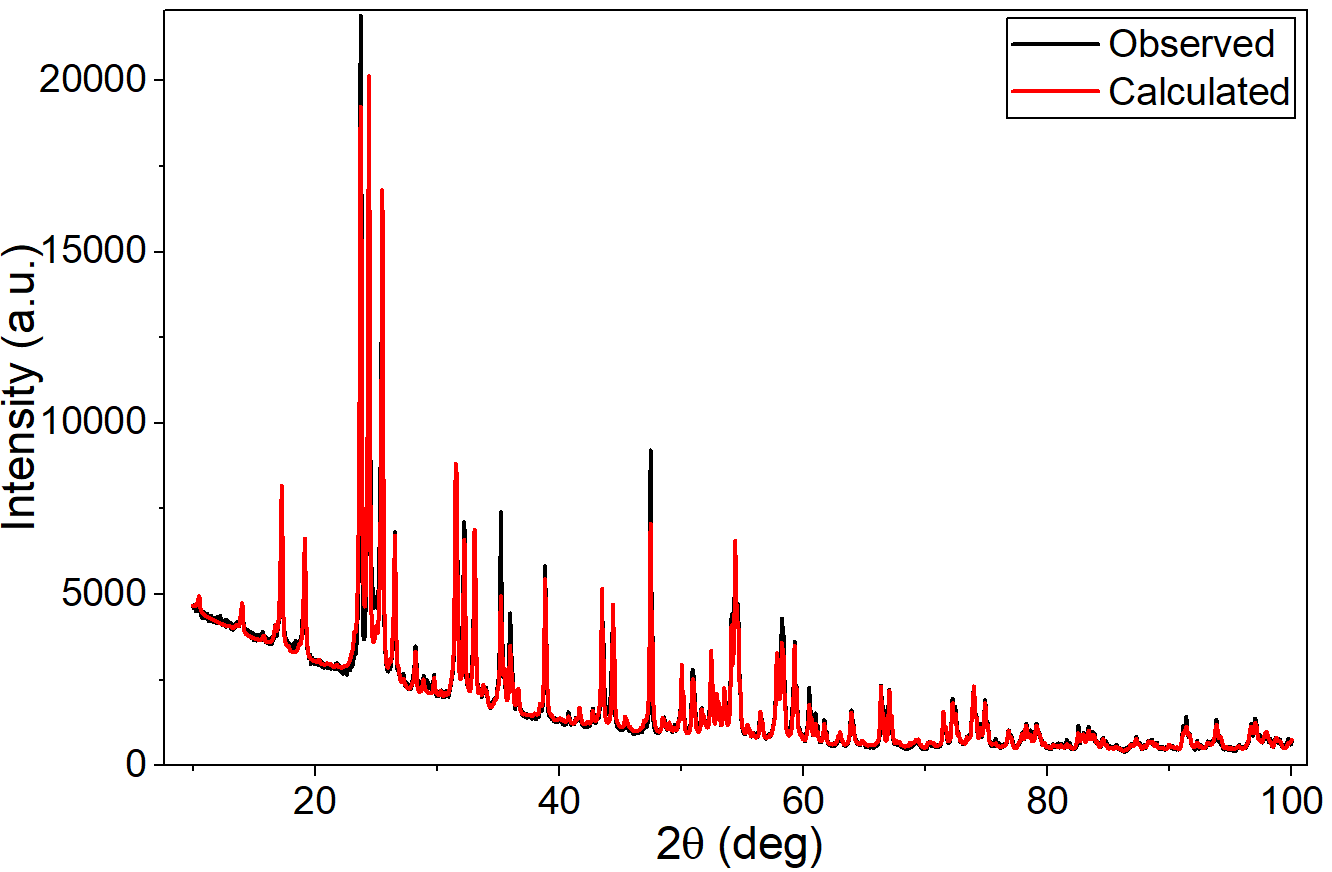}
    \caption{}
    \label{Nb25_XRD}
\end{subfigure}
\begin{subfigure}[b]{0.45\linewidth}
    \includegraphics[width=\linewidth]{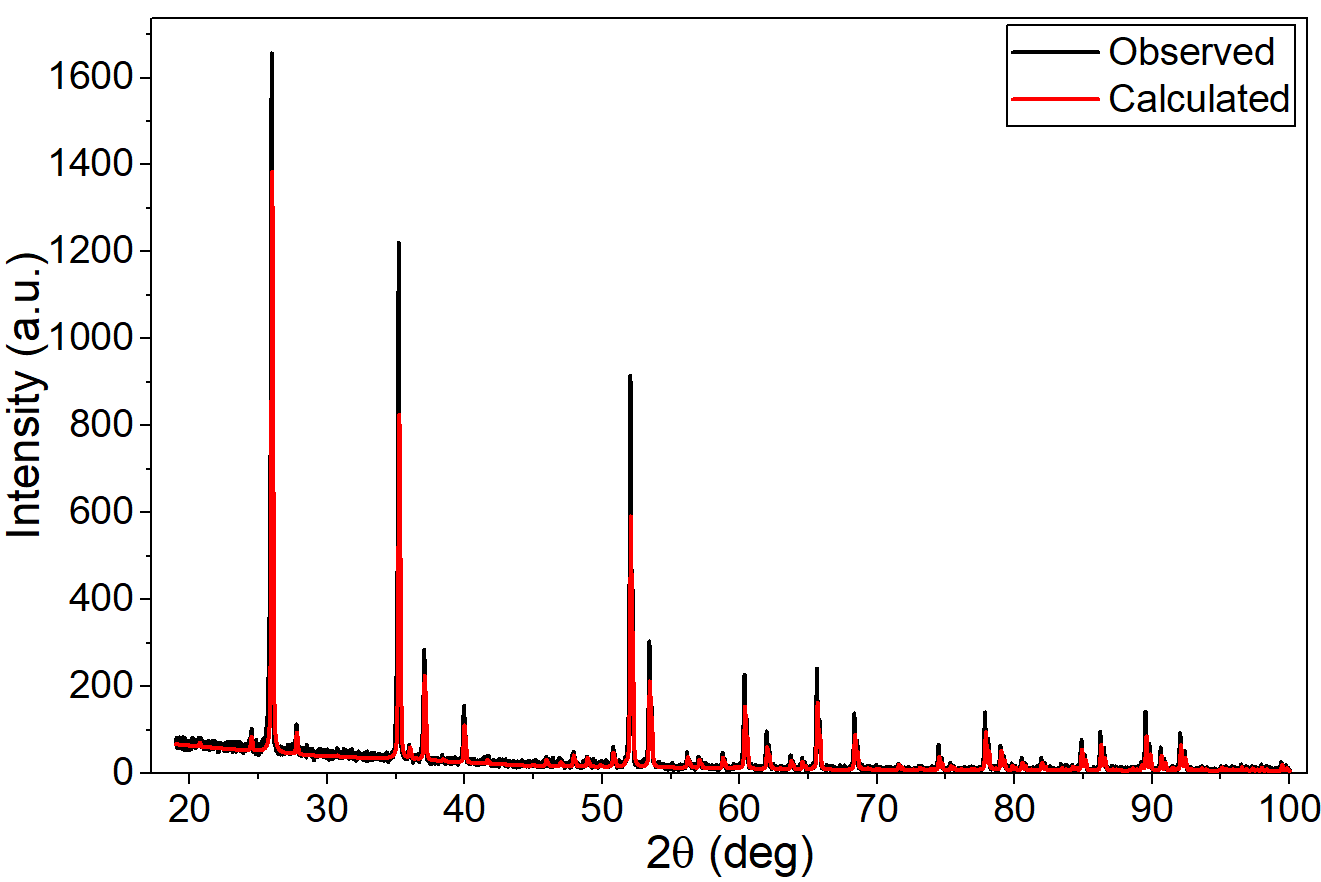}
    \caption{}
    \label{Nb02_XRD}
\end{subfigure}

\caption{Comparison of observed and calculated XRD patterns for (a) \ch{Nb2O5} and (b) \ch{NbO2}. The measured room‑temperature diffraction data (black) are overlaid with the corresponding reference patterns (red) obtained from phase‑matching analysis. The close agreement between observed and calculated peak positions and relative intensities confirms that both powders are microcrystalline and correspond to the (a) monoclinic \ch{Nb2O5} and (b) tetragonal \ch{NbO2} phases. No additional reflections are observed that may indicate the presence of additional phases.}
\label{XRD}
\end{figure}

To chemically characterize the \ch{Nb2O5} and \ch{NbO2} powders, we measured representative samples of each using XRD. Fig.\ref{XRD} shows the room‑temperature diffraction patterns for both oxides. The presence of sharp and well‑defined Bragg peaks indicates that each powder is predominantly microcrystalline. We analyzed the diffraction patterns using Jade software to determine the phase and assess the purity of each oxide type. The \ch{Nb2O5} pattern matches the monoclinic phase reported in Ref.\cite{Burdeze1969}. Likewise, the \ch{NbO2} pattern matches the tetragonal phase reported in Ref.\cite{sakata1979}.

In both cases, the measured peaks align with established reference datasets, and no additional reflections attributable to secondary phases or impurities are observed. This data provide strong evidence that the powders used in this study are microcrystalline and phase‑pure.

\section{Electromagnetic Simulations}

In order to model the amount of energy loss from various sources in our system, we utilize finite-element analysis to identify the appropriate electromagnetic properties for our setup and determine the electric field distribution required to calculate the participation ratio $F$ (see the definition in the main paper), as described in Eq.\ref{ratioF}.

\begin{equation}
\begin{array}{cc}
     F = \frac{U_{sample}} {U_{total}} = \frac{U_{sample}} {U_{cavity} + U_{substrate} + U_{sample}}  \\ \\
     U_{cavity} = \frac{1}{4}\int_{V_{cavity}}\varepsilon_{cavity}|\vec{E}|^2 dV \\ \\
     U_{substrate} = \frac{1}{4}\int_{V_{substrate}}\varepsilon_{substrate}|\vec{E}|^2 dV \\ \\
     U_{sample} = \frac{1}{4}\int_{V_{sample}}\varepsilon_{sample}|\vec{E}|^2 dV
      
\end{array}
\label{ratioF}
\end{equation}

Here, $U$ denotes the average energy of the electric field within a region, as specified by the subscript, $\varepsilon$ is the permittivity of the medium, $|\vec{E}|$ is the amplitude of the electric field at resonance at a given position, and $V$ is the volume that is being integrated over. 

\begin{figure}[H]
\centering
\begin{subfigure}[b]{0.45\linewidth}
    \includegraphics[width=\linewidth]{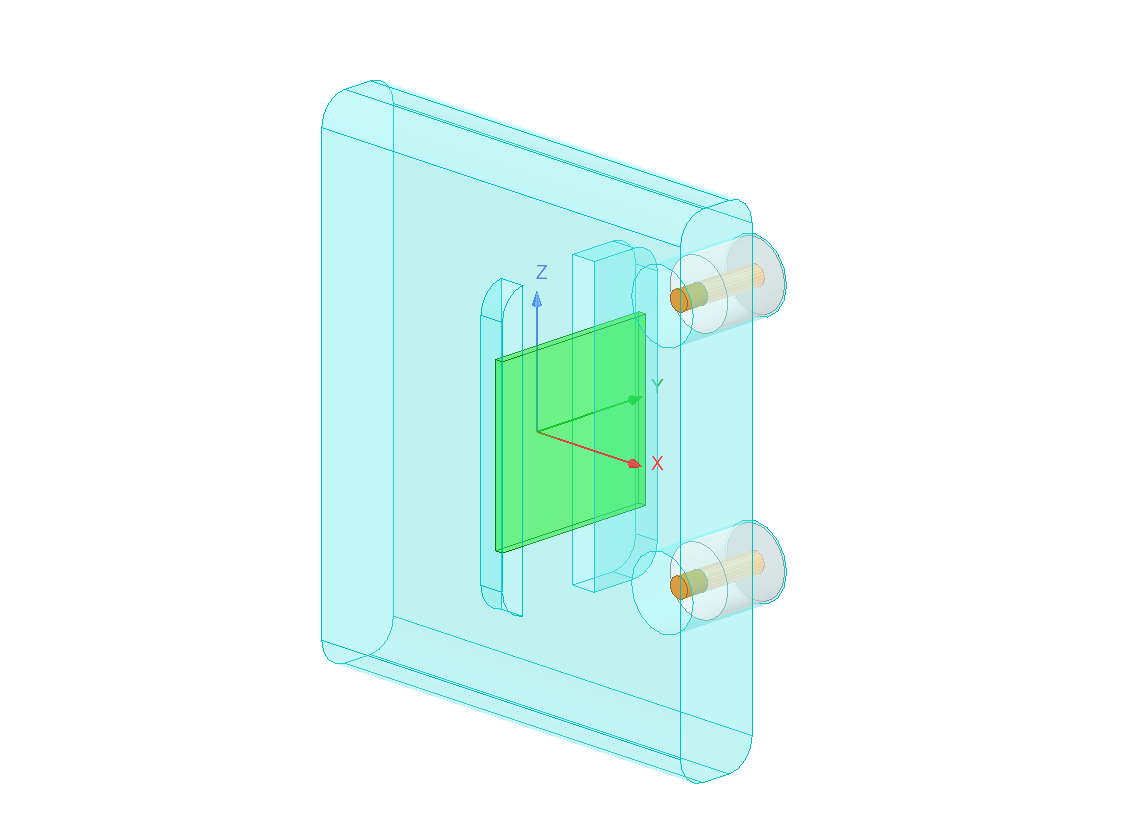}
    \caption{}
    \label{SubstrateOnly}
\end{subfigure}
\begin{subfigure}[b]{0.45\linewidth}
    \includegraphics[width=\linewidth]{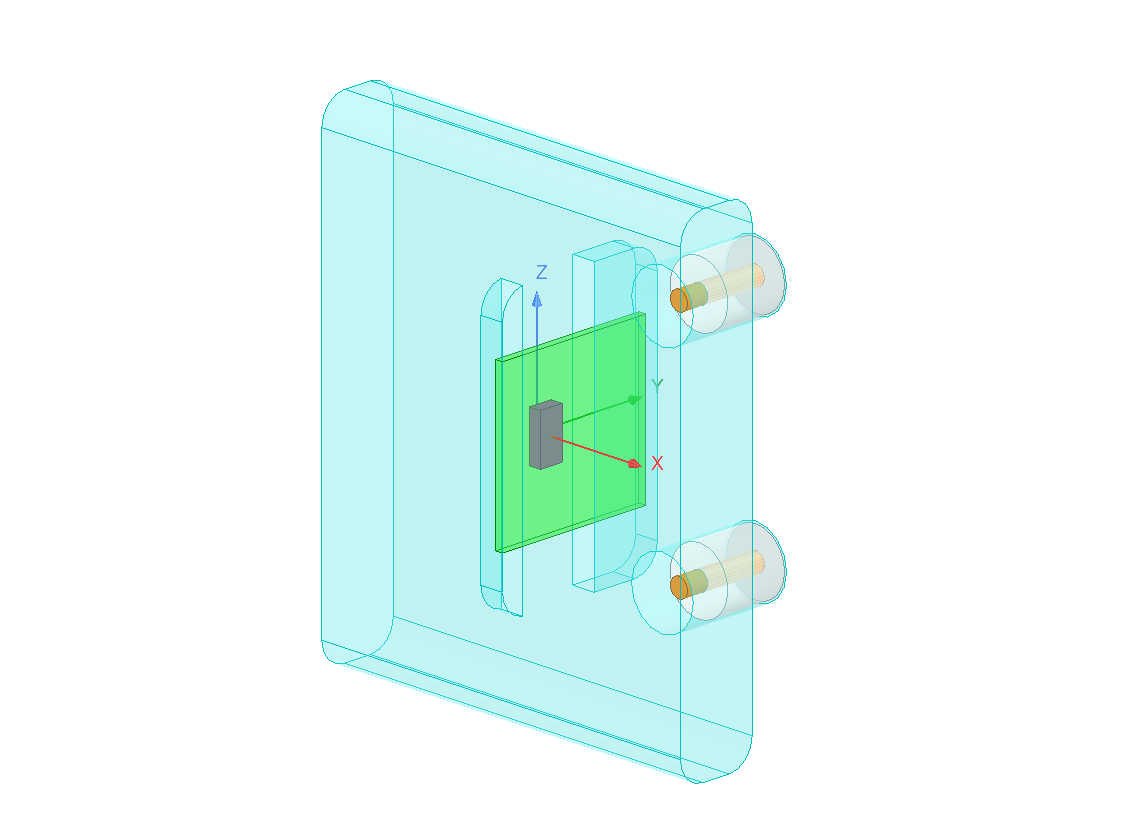}
    \caption{}
    \label{OxideSample}
\end{subfigure}

\caption{Geometry of the simulation for (a) the substrate-loaded cavity and (b) the substrate + the oxide-sample-loaded cavity. The blue area shows the empty space of the cavity, green shows the substrate, purple is the oxide sample, gold is the antennae from the SMA jack connectors, and white is the teflon casing surrounding the antennae. Cavity dimensions are $25\times5\times30$\,mm for the XYZ directions respectively.}
\label{SubstrateSetUp}
\end{figure}

The analysis was done in two main stages. The first stage modeled the cavity and bare substrate with no sample (Fig.\ref{SubstrateOnly}), with the assumption that $\varepsilon_{substrate} = 9.3$ \cite{krupka1999complex}, to optimize the electrical properties of the involved components. The driven modal solution yields the $S_{21}$ of the fundamental cavity mode, and this is used to match the measured $Q_\text{L}$ and $f_0$ of the sapphire-loaded cavity, by changing the cavity conductivity and the substrate's electric loss tangent. Once these parameters are established, the oxide is introduced in the simulation (Fig.\ref{OxideSample}), with the volume determined by the density of \ch{Nb2O5} ($4.6$ \text{$\frac{mg}{mm^3}$} \cite{shih2023investigation}) and the weight of the added oxide during sample preparation. It was assumed that $\varepsilon_{sample} = 45$ \cite{fuschillo1975dielectric}. At this point, the $S_{21}$ spectrum was simulated again, and the electrical loss tangent of \ch{Nb2O5} was tuned to match the measured $Q_\text{L}$ for TLS-hosting samples at all powers.

\begin{figure}[]
\centering
\begin{subfigure}[h]{0.49\linewidth}
    \includegraphics[width=\linewidth]{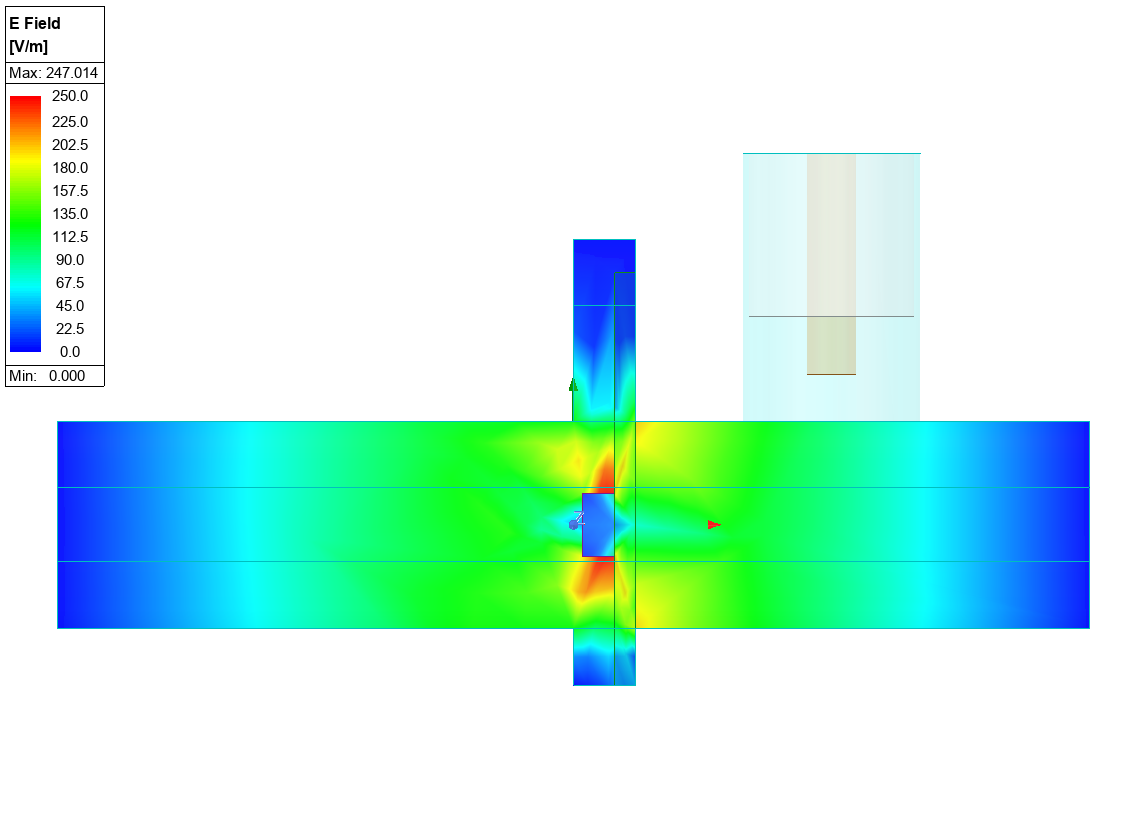}
    \caption{}
\end{subfigure}

\begin{subfigure}[h]{0.49\linewidth}
    \includegraphics[width=\linewidth]{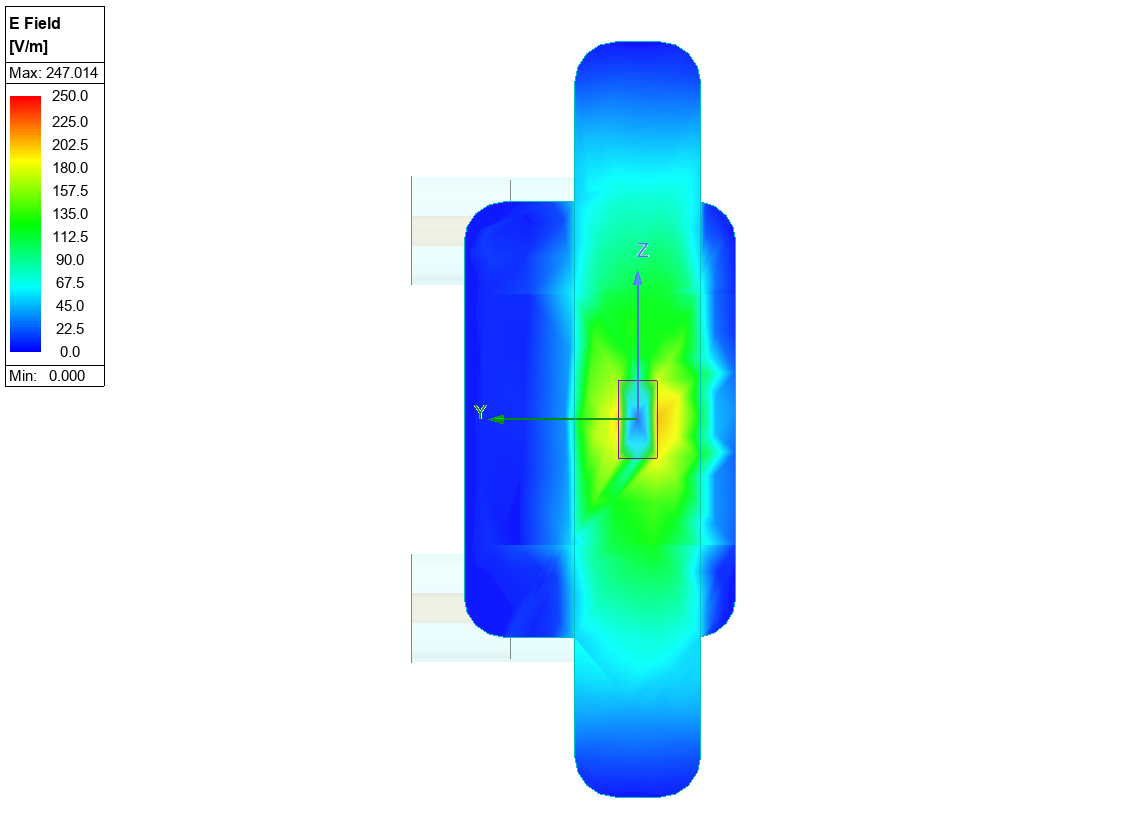}
    \caption{}
\end{subfigure}
\begin{subfigure}[h]{0.49\linewidth}
    \includegraphics[width=\linewidth]{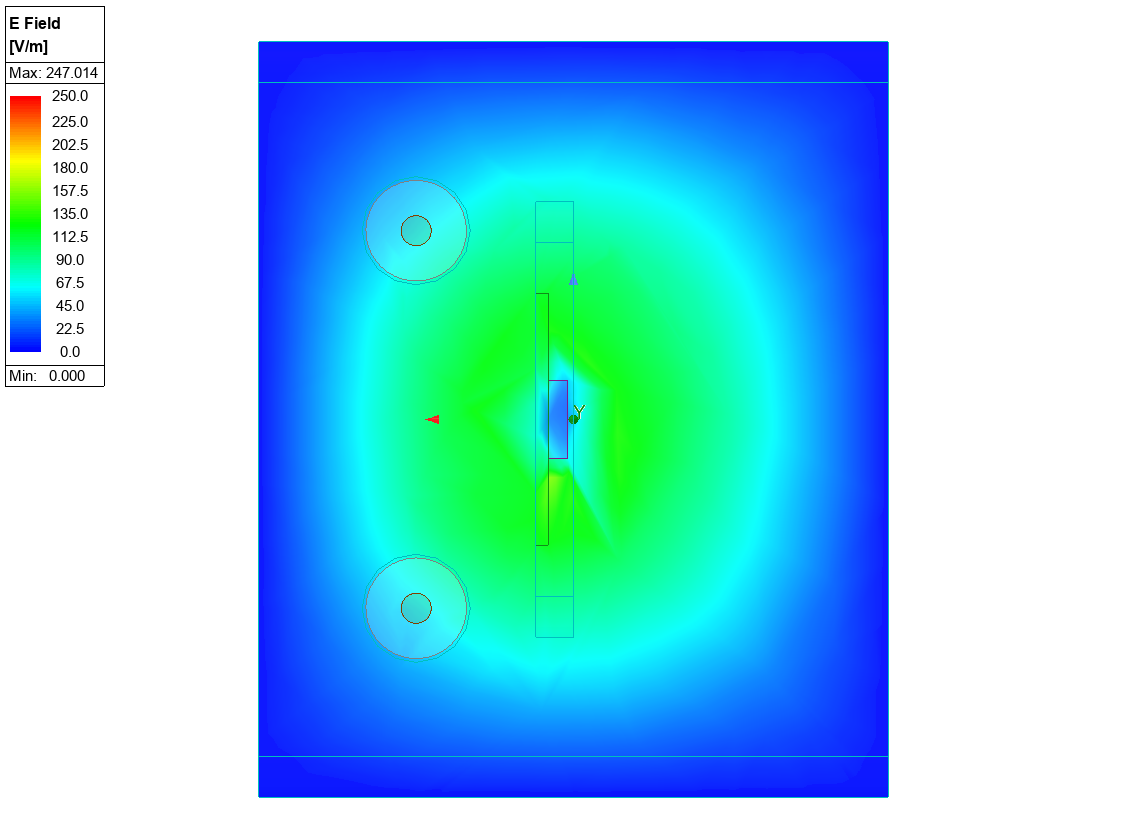}
    \caption{}
\end{subfigure}
\caption{Electric field distribution at resonance along the (a) XY, (b) YZ, and (c) XZ planes. Input power to the SMA jack is $-40$ dBm for this simulation.}
\label{EfieldDist}
\end{figure}

With the simulation established, we can determine the electric field distribution within all regions of the cavity, which is shown in the cross-sectional views in Fig.\ref{EfieldDist}. Mesh size is about $0.01$\text{\%} the total cavity volume. Note the found $F \approx 0.02$ with this method is many orders of magnitude higher than previously reported simulations (previous reports find F ranging from $10^{-3}$ to $10^{-9}$), with the reason attributed to the TLS hosting material occupying a much larger volume in this set-up compared to previous studies \cite{Romanenko2017, Verjauw2021}.

\end{document}